# RKKY interaction of magnetic impurities in node-line semimetals

Zhong-Yi Wang, Da-Yong Liu and Liang-Jian Zou

*Key Laboratory of Materials Physics, Institute of Solid State Physics, HFIPS, Chinese Academy of Sciences, P. O. Box 1129, Hefei 230031, China*

*and Science Island Branch of Graduate School, University of Science and Technology of China, Hefei 230026, China*

**Abstract**

Motivated by the recent upsurge in research of three-dimensional topological semimetals (SMs), we theoretically study the RKKY interaction between magnetic impurities in node-line SMs with and without the chirality and obtain the analytical expressions. We find that unique toroidal Fermi surface (FS) in nodal-line SMs, distinctly different from the spheroid FS in the SMs with the point nodes, has significant influences on the RKKY interaction, leading to strong anisotropic oscillation and unique decay features. In the direction perpendicular to node-line plane, as usual, there is only one oscillation period related to the Fermi energy. In contrast, in the node-line plane, the RKKY interaction form a beating pattern and oscillates more rapidly with two distinct periods: one is coming from the Fermi energy and the other is from the radius of node-line. More importantly, inside nodal-line SMs bulk, the decay rate of RKKY interaction manifests a typical two-dimensional feature for impurities aligned along the direction perpendicular to nodal-line plane. Furthermore, the magnetic interactions in nodal-line SMs with linear and quadratic dispersions in the nodal-line plane are compared. We also discuss the possible application of the present theory on realistic NLSM ZrSiSe. Our results shed the light for application of magnetically doped node-line SMs in spintronics.

**PACS no:** 75.30.Hx, 71.55.Ak, 03.65.Vf, 75.10.-b

\* **Correspondence author:** zou@theory.issp.ac.cn

## I. Introduction

The discovery of quantum anomalous Hall effect [1] in magnetically doped topological insulators [2] stirs great interesting on magnetic interaction in topological materials [3-5], specially the indirect magnetic coupling between doped local magnetic impurities. As an indirect exchange coupling of two localized magnetic impurities, the Ruderman-Kittel-Kasuya-Yosida (RKKY) interaction [6-8], which plays key roles in the magnetism of Kondo lattices [9], heavy fermions [10], diluted magnetic metals and diluted magnetic semiconductor [11,12], is regarded as a crucial role for magnetic ordering of the impurities in topological insulators [13]. Actually, considering the fact that the realization of the RKKY coupling is controlled by itinerant electrons, the recently discovered topological semimetals (SMs), including Dirac SMs [14-18], Weyl SMs [19, 20] and node-line SMs [21-27], naturally serves as a good platform to explore more novel RKKY interaction properties due to their unique electronic band structures.

In the Weyl and Dirac SMs, the conduction and valence bands contact at discrete points, which are called Weyl or Dirac nodes. The quasiparticles near these nodes in the Brillouin zone follow the relativistic Dirac or Weyl equation and disperse linearly in all directions [27]. As a direct consequence of the unique dispersion, the RKKY interaction in doped graphene [28-30] decays as $1/R^3$ rather than $1/R^2$ in conventional 2D materials [31], which is attributed to the chiral nature of relativistic electrons in graphene [28]. Similarly, in the three-dimensional (3D) Dirac/ Weyl SMs, Chang *et al.* [32] found that the internode process, as well as the unique 3D spin-momentum locking, has significant influences on the RKKY interaction, resulting in both a Heisenberg and an Ising term, and an additional Dzyaloshinsky-Moriya term if the inversion symmetry is absent. Furthermore, in the Weyl SMs, due to a pair of Weyl points separated in the momentum space, Hosseini *et al.* [33] found a new spin-frustrated term and a beating pattern oscillation behavior depending on the direction between two magnetic impurities. For the case of magnetic impurities placed near the surface of the Weyl SMs, Duan *et al.* [34] reported that the interplay between Fermi arc and bulk states near the boundary contributes the RKKY interaction with magnitude comparable with or exceeding the bulk contribution. These studies disclosed the major unique characters of the RKKY interaction in topological Dirac and Weyl SMs.

Although the RKKY interaction in topological SMs has been extensively studied, most works focused on the Dirac SMs and Weyl SMs, and the situation in node-line SMs (NLSMs) remain much less explored. Different from Dirac and Weyl SMs, the conduction and valence bands in NLSMs touch along ring-shaped line node and disperse linearly in direction perpendicular to nodal ring [27], which is realized in ZrSiSe [21, 25-26], PbTaSe$_2$ [22], and black phosphorus under pressure [41]. When the system is doped, the FS of NLSMs has a toroidal shape in momentum space [35,36]. These unique band characters of NLSMs will lead to distinct indirect magnetic coupling.

In this work we present a systematic study on the RKKY interaction in NLSMs, and show that special band structure leads to strong anisotropic oscillation and decay features of RKKY interaction: the RKKY interaction displays a beating behavior with two distinct periods related to the Fermi energy and the radius of node line for the impurities deposited in the nodal-line plane. In contrast, for the impurities aligned along the direction perpendicular to nodal-line plane, beating behavior degenerate to a conventional oscillation with period related to the Fermi energy. And we demonstrate that NLSMs have unique decay features in these two representative directions, considerably different from the $1/R^5$ ($1/R^3$) decaying law [32,33] for zero (finite) Fermi energy in the Dirac/Weyl SMs. The most prominent is a typical two-dimensional decay feature for impurities embedded deeply inside nodal-line SMs bulk and aligned along the direction perpendicular to nodal-line plane. We apply our results to the realistic NLSM compound ZrSiSe.

The paper is organized as follows. In Sec. II, we outline the model Hamiltonian and derive general forms of RKKY range functions. In Sec. III, we present approximately analytical results and exact numerical results for two most representative alignments of impurities, one along the direction perpendicular to nodal ring and another in the nodal-line plane; also we analyze the possible application in realistic NLSM compound. Finally, we compare the RKKY couplings in the NLSMs with linear and quadratic dispersions in the nodal-line plane, and draw a brief conclusion in Sec. IV.

**II. Model Hamiltonian and Theoretical Formulas**

We consider the following Hamiltonian to describe the low-energy physics of NLSMs possessing linear dispersion [35] in the nodal-line plane and the chirality,

$$H_0 = \chi V_F \big[ (k_\rho - k_0)\sigma_x + k_z \sigma_y \big] \qquad (1)$$

where $k_\rho = \sqrt{k_x^2 + k_y^2}$, $V_F$ is the Fermi velocity in the z direction, $k_0$ is the radius of the nodal ring in the xy plane; $\chi = \pm$ denote two kinds of chirality; Pauli matrices $\sigma_{x,y}$ denote to the real spin degree of freedom of electrons. We define the $k_z$ direction of linearly dispersion as the vertical direction, and the nodal-line plane perpendicular to the $k_z$ as the transverse plane.

The s-d exchange interaction between the itinerant electrons and magnetic impurity $S_i$ located at $R_i$ can be expressed as [32,37]

$$H_i = (J\tau_0 + \lambda\tau_x)\vec{S}_i \cdot \vec{\sigma}\delta(r - R_i) \tag{2}$$

where J and $\lambda$ represent the strengths of the s-d exchange interaction in progresses between the same chirality and between different chirality, respectively. Note that J and $\lambda$ has the dimensionality: (energy)×(length)$^D$, D is the dimension of the system [38]. The Pauli matrices $\tau_x$ and the identity $\tau_0$ matrix act on the chirality space [32]. We assume that the coupling constants J and $\lambda$ are small so that the s-d interaction (2) can be treated as a perturbation to the node-line Hamiltonian (1). According to the second order perturbation theory, at zero temperature the RKKY interaction between two magnetic impurities can be expressed as [32, 37-38]

$$H_{1,2}^{RKKY} = \sum_{\chi,\chi',\alpha,\beta}[J^2\delta_{\chi\chi'} + \lambda^2(1-\delta_{\chi\chi'})]S_1^\alpha S_2^\beta Im\{-\frac{1}{\pi}\int_{-\infty}^{\varepsilon_F}d\varepsilon\,Tr[\sigma_\alpha G_\chi(\vec{R};\varepsilon)\sigma_\beta G_{\chi'}(-\vec{R};\varepsilon)]\} \tag{3}$$

where $\varepsilon_F$ is the Fermi energy, $\vec{R} = \vec{R_1} - \vec{R_2}$, Tr means a trace over the spin and pseudospin degree of freedom, $G_\chi(\vec{R};\varepsilon)$ is the Green's function of the free electrons in the node-line SMs in the energy-coordinate representation.

Since Green's function in momentum space takes the form $G_\chi^{-1}(\vec{k},\varepsilon) = (\varepsilon + i\eta)\sigma_0 - H_0$, so $G_\chi(\vec{R};\varepsilon)$ can be obtained from $G_\chi^{-1}(\vec{k},\varepsilon)$ through performing a Fourier transformation,

$$G_\chi(\vec{R};\varepsilon) = \int \frac{d^3\vec{k}}{(2\pi)^3} \frac{\varepsilon + H_0}{(\varepsilon+i\eta)^2 - E_k^2} e^{i\vec{k}\cdot\vec{R}}, \tag{4}$$

where $E_k = V_F\sqrt{(k_\rho - k_0)^2 + k_z^2}$. Under the cylindrical coordinate, i.e. $\vec{R} = (\rho, \varphi, z)$, Eq. (4) can be written as

$$G_\chi(\vec{R};\varepsilon) = G_0(\vec{R},\varepsilon)\sigma_0 + G_1(\vec{R},\varepsilon)\chi\sigma_x + G_2(\vec{R},\varepsilon)\chi\sigma_y, \tag{5}$$

with the partial Green's functions

$$G_0(\vec{R},\varepsilon) = \frac{2\varepsilon}{(2\pi)^2}\left(\int_0^{k_0}dS\int_0^\pi d\phi + \int_{k_0}^\infty dS\int_0^{arccos(-k_0/S)}d\phi\right)\frac{J_0[(S\cos\phi+k_0)\rho]\cos(S\sin\phi\cdot z)}{(\varepsilon+i\eta)^2 - V_F^2 S^2}S(S\cos\phi + k_0),$$

$$G_1(\vec{R},\varepsilon) = \frac{2V_F}{(2\pi)^2}\left(\int_0^{k_0}dS\int_0^\pi d\phi + \int_{k_0}^\infty dS\int_0^{arccos(-k_0/S)}d\phi\right)\frac{J_0[(S\cos\phi+k_0)\rho]\cos(S\sin\phi\cdot z)}{(\varepsilon+i\eta)^2 - V_F^2 S^2}S^2\cos\phi\,(S\cos\phi + k_0),$$

$$G_2(\vec{R},\varepsilon) = i\frac{2V_F}{(2\pi)^2}\left(\int_0^{k_0} dS \int_0^{\pi} d\phi + \int_{k_0}^{\infty} dS \int_0^{\arccos(-k_0/S)} d\phi\right) \frac{J_0[(S\cos\phi+k_0)\rho]\sin(S\sin\phi\cdot z)}{(\varepsilon+i\eta)^2 - V_F^2 S^2} S^2 \sin\phi (S\cos\phi+k_0). \tag{6}$$

where $J_0$ is the zeroth-order Bessel function.

Using the relations $G_{0,1}(-\vec{R},\varepsilon) = G_{0,1}(\vec{R},\varepsilon)$ and $G_2(-\vec{R},\varepsilon) = -G_2(\vec{R},\varepsilon)$, we have

$$G_\chi(-\vec{R};\varepsilon) = G_0(\vec{R},\varepsilon)\sigma_0 + G_1(\vec{R},\varepsilon)\chi\sigma_x - G_2(\vec{R},\varepsilon)\chi\sigma_y. \tag{7}$$

Substituting Eqs. (4)-(7) into Eq. (3), we obtain the RKKY interaction of node-line SMs

$$H_{RKKY,\chi} = [\lambda^2 F_1(R) + J^2 F_2(R)]\vec{S}_1\cdot\vec{S}_2 - 2(J^2-\lambda^2)F_3(R)S_1^y S_2^y + 2(J^2-\lambda^2)F_4(R)S_1^x S_2^x, \tag{8}$$

where the range functions are given as

$$F_1(R) = f_0(R) + f_1(R) - f_2(R),$$

$$F_2(R) = f_0(R) - f_1(R) + f_2(R),$$

$$F_3(R) = f_2(R), \quad F_4(R) = f_1(R), \tag{9.1}$$

with

$$f_i(R) = -\frac{4}{\pi} \text{Im} \int_{-\infty}^{\varepsilon_F} d\varepsilon\, G_i(\vec{R},\varepsilon)^2, i=0,1,2. \tag{9.2}$$

On the other hand, in the node-line SMs without the chirality, the Hamiltonian has following form $H_0 = V_F[(k_\rho - k_0)\tilde{\sigma}_x + k_z\tilde{\sigma}_y]\otimes\sigma_0$, where the Pauli matrices $\tilde{\sigma}_{x,y}$ and $\sigma_0$ refer to pseudospin and real spin degrees of freedom, respectively. Since there is no progress between the different chirality, the s-d exchange interaction takes the form $H_i = J\vec{S}_i\cdot\vec{\sigma}\delta(r-R_i)$. Using a similar approach, we get the corresponding RKKY coupling

$$H_{RKKY} = J^2 F_1(R)\vec{S}_1\cdot\vec{S}_2. \tag{10}$$

To display the spatial dependence of the magnetic coupling, which reduces to the classic form, we have to calculate the integrals ($\int_{-\infty}^{\varepsilon_F} d\varepsilon\, G_i(\vec{R},\varepsilon)^2$) in Eq. (9.2). Since the $G_i(\vec{R},\varepsilon)$ are complicated and could not be integrated analytically. But when the size of the nodal ring is so large [35] that $V_F k_0 \gg |\varepsilon|$, we obtain the following compact approximately analytical forms

$$g_0(\vec{R},\varepsilon) = \frac{2k_0\varepsilon}{(2\pi)^2}\int_0^\infty dS \int_0^\pi d\phi\, \frac{J_0[(S\cos\phi+k_0)\rho]\cos(S\sin\phi\cdot z)}{(\varepsilon+i\eta)^2 - V_F^2 S^2} S,$$

$$g_1(\vec{R},\varepsilon) = \frac{2V_F k_0}{(2\pi)^2}\int_0^\infty dS \int_0^\pi d\phi\, \frac{J_0[(S\cos\phi+k_0)\rho]\cos(S\sin\phi\cdot z)}{(\varepsilon+i\eta)^2 - V_F^2 S^2} S^2 \cos\phi,$$

$$g_2(\vec{R},\varepsilon) = i\frac{2V_F k_0}{(2\pi)^2}\int_0^\infty dS \int_0^\pi d\phi\, \frac{J_0[(S\cos\phi+k_0)\rho]\sin(S\sin\phi\cdot z)}{(\varepsilon+i\eta)^2 - V_F^2 S^2} S^2 \sin\phi. \tag{11}$$

It is considering that the $G_i(\vec{R},\varepsilon)$ has less contribution [29,37] to integrals ($\int_{-\infty}^{\varepsilon_F} d\varepsilon\, G_i(\vec{R},\varepsilon)^2$) in the deep energy area ($\varepsilon \ll 0$). Thus within the large ring approximation ($V_F k_0 \gg |\varepsilon|$), for the topology state of node-line SMs ($|\varepsilon_F| \ll V_F k_0$), the $G_i(\vec{R},\varepsilon)$ can be replaced by the $g_i(\vec{R},\varepsilon)$ in integrals($\int_{-\infty}^{\varepsilon_F} d\varepsilon\, G_i(\vec{R},\varepsilon)^2$). As a result, we figure out approximately analytical solutions in good agreement with numerical solutions. In view of the anisotropy of the dispersion, we focus on two representative alignments of impurities: one is along the vertical direction, another is in the transverse plane, as discussed at length in what follows.

## III. RESULTS AND DISCUSSIONS

### A.  Impurities aligned along the vertical direction

When the two magnetic impurities are deposited along the vertical direction, i.e., $\rho = 0$, $J_0(0) = 1$, the $g_i(\vec{R},\varepsilon)$ reduces to a simple form

$$g_0^z(z,\varepsilon) = \frac{2k_0 \varepsilon}{(2\pi)^2} \int_0^\infty dS \int_0^\pi d\phi\, \frac{\cos(S\sin\phi \cdot z)}{(\varepsilon+i\eta)^2 - V_F^2 S^2} S,$$

$$g_1^z(z,\varepsilon) = 0,$$

$$g_2^z(z,\varepsilon) = i\frac{2V_F k_0}{(2\pi)^2} \int_0^\infty dS \int_0^\pi d\phi\, \frac{\sin(S\sin\phi \cdot z)}{(\varepsilon+i\eta)^2 - V_F^2 S^2} S^2 \sin\phi = -i\frac{V_F}{\varepsilon}\frac{\partial}{\partial z} g_0^z(z,\varepsilon). \quad (12)$$

Although the $g_i^z(z,\varepsilon)$ can not be integrated analytically, by using the heuristic approach as described by Y. Sun $et\ al$ [37], we can obtain the asymptotic behavior

$$g_0^z(z,\varepsilon) \approx -i\frac{k_0}{4V_F}\sqrt{\frac{2}{\pi}}\sqrt{\frac{\varepsilon}{V_F z}} e^{i\left(\frac{\varepsilon}{V_F} z - \frac{\pi}{4}\right)}, \quad (13)$$

$$g_2^z(z,\varepsilon) \approx g_0^z(z,\varepsilon)\left(i\frac{V_F}{\varepsilon}\frac{1}{2z} + 1\right). \quad (14)$$

After some algebras, we obtain the each term ($f_i(R)$) of the range function in Eq. (9.2)

$$f_0^z(z) = -\frac{k_0^2}{8\pi^2 V_F}\frac{1}{z^3}[\cos(2k_F z) + 2k_F z \sin(2k_F z)],$$

$$f_1^z(z) = 0,$$

$$f_2^z(z) = -\frac{k_0^2}{8\pi^2 V_F}\frac{1}{z^3}[3\cos(2k_F z) + 2k_F z \sin(2k_F z)]. \quad (15)$$

with $k_F = \varepsilon_F/V_F$. Accordingly, the RKKY interactions in node-line SMs with the chirality is

$$H^z_{RKKY,\chi} = [\lambda^2 F_1(z) + J^2 F_2(z)](S_1^x S_2^x + S_1^z S_2^z) + [\lambda^2 F_2(z) + J^2 F_1(z)]S_1^y S_2^y, \tag{16.1}$$

where the range functions $F_{1,2}(z)$ are given by

$$F_1(z) = \frac{k_0^2}{4\pi^2 V_F z^3}\cos(2k_F z),$$

$$F_2(z) = -\frac{k_0^2}{4\pi^2 V_F z^3}[2\cos(2k_F z) + 2k_F z \sin(2k_F z)]. \tag{16.2}$$

To verify the accuracy of the analytical results Eq. (16), we compare numerical results with analytical results of RKKY range functions $F_1(z)$ and $F_2(z)$ in Fig. 1 for two magnetic impurities aligned along the vertical direction. We find that our approximately analytical solutions agree well with numerical solutions.

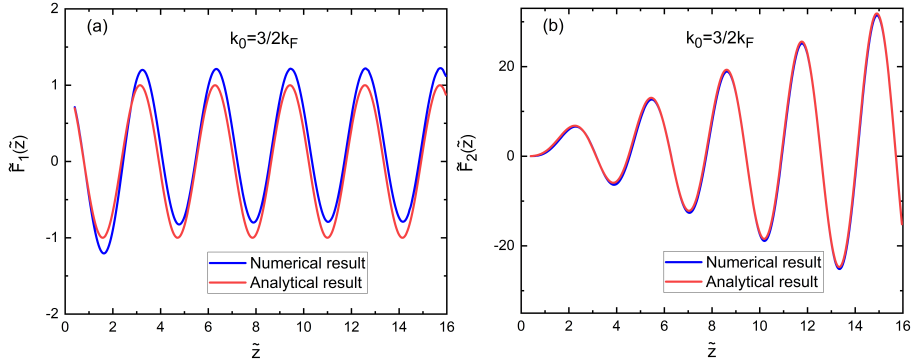

Fig. 1 The range functions of the RKKY interaction of node-line SMs as a function of the dimensionless parameter $\tilde{z}$ for two impurities aligned along the vertical direction. Here $k_0 = 3/2k_F$. (a) and (b) are range functions $\tilde{F}_1(\tilde{z})$ and $\tilde{F}_2(\tilde{z})$ respectively, with $\tilde{F}_{1,2}(\tilde{z}) = \tilde{z}^3 F_{1,2}(\tilde{z})/C$, $\tilde{z} = k_F z$, $C = k_0^2 k_F^3/4\pi^2 V_F$.

Furthermore, at long range ($k_F z \gg 1$) and for finite doping, the RKKY interaction with the chirality reduces to a simple form

$$H^z_{RKKY,\chi} \approx \frac{k_0^2 k_F \sin(2k_F z)}{2\pi^2 V_F z^2}\left[-J^2 \vec{S}_1 \cdot \vec{S}_2 + (J^2 - \lambda^2)S_1^y S_2^y\right]. \tag{17}$$

It is interesting to note that, compared with the $\cos(2k_F R)/R^3$ dependence of RKKY in conventional three-dimensional metal, Dirac SMs and Weyl SMs, the present $\sin(2k_F z)/z^2$ dependence is the remarkable two-dimensional characteristic [31]. This stems from the distinct Fermi surface (FS) structure of node-line SMs. Since the toroidal FS of node-line SMs can be decomposed into infinitely equivalent two-dimensional annulus FSs by intersecting the nodal ring in the vertical direction, and each annulus FS distributes in a two-dimensional plane in momentum space [35]. The full RKKY interaction in the vertical direction is contributed by the sum of all such two-dimensional FS, leading to significant two-dimensional characteristic.

As a comparison, in the node-line SMs without the chirality, we obtain the RKKY interaction of two impurities spins

$$H_{RKKY}^z = \frac{J^2 k_0^2}{4\pi^2 V_F z^3} \cos(2k_F z) \vec{S}_1 \cdot \vec{S}_2. \tag{18}$$

Distinctly different from the case with the chirality, here the RKKY interaction displays $1/z^3$ decaying behavior, which is a direct result of the pseudospin-momentum locking [28] in the vertical direction. More importantly, the amplitude of RKKY interaction is independent of Fermi energy, which means the variation of Fermi energy just change the oscillation period.

For the intrinsic case ($k_F = 0$), the RKKY interaction becomes nonoscillatory:

$$H_{RKKY,\chi}^z = \frac{k_0^2}{4\pi^2 V_F z^3}[(\lambda^2 - 2J^2)\vec{S}_1 \cdot \vec{S}_2 + (3J^2 - 3\lambda^2)S_1^y S_2^y],$$

$$H_{RKKY}^z = \frac{J^2 k_0^2}{4\pi^2 V_F z^3} \vec{S}_1 \cdot \vec{S}_2. \tag{19}$$

In the case with the chirality, we can see that the ferromagnetism (FM) or antiferromagnetism (AFM) of magnetic impurities depends on the relative magnitude of $\lambda$ and $J$. But in the case without the chirality, exchange coupling between magnetic impurities is always antiferromagnetic. In both cases, it is obvious that their spatial dependences display as $1/z^3$, much slowly than the $1/z^5$ law in the intrinsic 3D Dirac and Weyl SMs [32-33], which is similar to the case in graphene. In fact, for the node-line SMs without the chirality, the RKKY interaction fall off as $1/z^3$ for both zero Fermi energy and finite Fermi energy in the vertical direction, displaying a typical behavior in graphene [28-30].

**B. Impurities aligned in the transverse plane**

Next, we consider the indirect magnetic coupling between two impurities deposited in the transverse plane ($z = 0$). In this situation, the $g_i(\vec{R}, \varepsilon)$ reduces to a simple form

$$g_0^\rho(\rho, \varepsilon) = \frac{2k_0 \varepsilon}{(2\pi)^2} \int_0^\infty dS \int_0^\pi d\phi \frac{J_0[(S\cos\phi + k_0)\rho]}{(\varepsilon + i\eta)^2 - V_F^2 S^2} S,$$

$$g_1^\rho(\rho, \varepsilon) = \frac{2V_F k_0}{(2\pi)^2} \int_0^\infty dS \int_0^\pi d\phi \frac{J_0[(S\cos\phi + k_0)\rho]}{(\varepsilon + i\eta)^2 - V_F^2 S^2} S^2 \cos\phi,$$

$$g_2^\rho(\rho, \varepsilon) = 0. \tag{20}$$

Considering the case of large ring approximation ($V_F k_0 \gg |\varepsilon|$), at large distance $k_0 \rho \gg 1$, $g_0^\rho(\rho, \varepsilon)$ can be simplified to

$$g_0^\rho(\rho,\varepsilon) \approx \frac{2k_0\varepsilon}{(2\pi)^2}\sqrt{\frac{2}{\pi k_0 \rho}} \int_0^\infty dS \int_0^\pi d\phi \frac{\cos\left[(S\cos\phi + k_0)\rho - \frac{\pi}{4}\right]S}{(\varepsilon + i\eta)^2 - V_F^2 S^2} = \sqrt{\frac{2}{\pi k_0 \rho}} \cos\left(k_0\rho - \frac{\pi}{4}\right) g_0^z(\rho,\varepsilon). \quad (21)$$

Repeating the above method, we obtain

$$g_1^\rho(\rho,\varepsilon) \approx i\sqrt{\frac{2}{\pi k_0 \rho}} \sin\left(k_0\rho - \frac{\pi}{4}\right) g_2^z(\rho,\varepsilon). \quad (22)$$

Thus we can easily obtain the relative terms ($f_i(R)$) of the range function in Eq. (9.2)

$$f_0^\rho(\rho) = \frac{2}{\pi k_0 \rho} \cos^2\left(k_0\rho - \frac{\pi}{4}\right) f_0^z(\rho),$$

$$f_1^\rho(\rho) = -\frac{2}{\pi k_0 \rho} \sin^2\left(k_0\rho - \frac{\pi}{4}\right) f_2^z(\rho),$$

$$f_2^\rho(\rho) = 0. \quad (23)$$

where the expression of $f_0^z(\rho)$ and $f_2^z(\rho)$ is from Eq. (15). Accordingly, the RKKY interactions with the chirality for two impurities aligned in the transverse plane is

$$H_{RKKY,\chi}^\rho = [\lambda^2 F_1(\rho) + J^2 F_2(\rho)](S_1^y S_2^y + S_1^z S_2^z) + [\lambda^2 F_2(\rho) + J^2 F_1(\rho)] S_1^x S_2^x, \quad (24.1)$$

where the range functions $F_{1,2}(\rho)$ are given by

$$F_1(\rho) = \frac{2}{\pi k_0 \rho}\left[\cos^2\left(k_0\rho - \frac{\pi}{4}\right) f_0^z(\rho) - \sin^2\left(k_0\rho - \frac{\pi}{4}\right) f_2^z(\rho)\right],$$

$$F_2(\rho) = \frac{2}{\pi k_0 \rho}\left[\cos^2\left(k_0\rho - \frac{\pi}{4}\right) f_0^z(\rho) + \sin^2\left(k_0\rho - \frac{\pi}{4}\right) f_2^z(\rho)\right].$$

Fig. 2 shows the comparison between the numerical results and analytical asymptotic ones of RKKY range functions $F_{1,2}(\rho)$ when two impurities are aligned in the transverse plane, we can see approximately analytical results according well with the numerical ones.

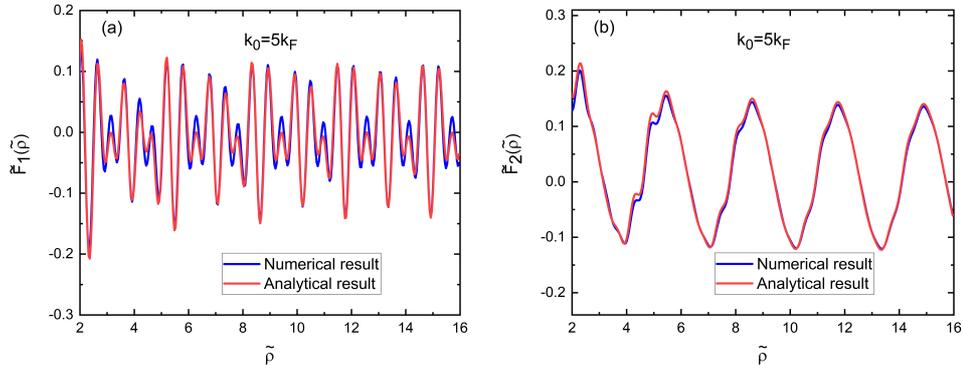

Fig. 2 The range function of the RKKY interaction of node-line SMs as a function of the dimensionless parameter $\tilde{\rho}$ for two impurities in the transverse plane. Here $k_0 = 5k_F$ (a), (b) are range functions $\tilde{F}_1(\tilde{\rho})$, $\tilde{F}_2(\tilde{\rho})$, respectively, with $\tilde{F}_{1,2}(\tilde{\rho}) = \tilde{\rho}^3 F_{1,2}(\tilde{\rho})/C$, $\tilde{\rho} = k_F\rho$, $C = k_0{}^2 k_F{}^3/4\pi^2 V_F$.

At long range ($k_F\rho \gg 1$) and for finite doping, the RKKY interaction with and without the chirality for impurities in the transverse plane reduces to simple forms

$$H_{RKKY,\chi}^{\rho} \approx -\frac{k_0 k_F \sin(2k_F\rho)}{2\pi^3 V_F \rho^3}\left\{[sin(2k_0\rho)\lambda^2 + J^2]\vec{S}_1 \cdot \vec{S}_2 - 2(J^2 - \lambda^2)\sin^2\left(k_0\rho - \frac{\pi}{4}\right)S_1^x S_2^x\right\},$$

$$H_{RKKY}^{\rho} \approx -\frac{J^2 k_0 k_F \sin(2k_F\rho)\sin(2k_0\rho)}{2\pi^3 V_F \rho^3}\vec{S}_1 \cdot \vec{S}_2. \tag{25}$$

In the case of $\lambda = J$, the Ising term in the node-line SMs with chirality vanishes. One could clearly see that in both cases, the RKKY interaction oscillate with two distinct periods: one is related to the reciprocal of radius of node-line in momentum space, $\pi/k_0$, another is related to the Fermi wavelength $\pi/k_F$. Thus we can observe a beating pattern in Fig. 2. For the topology state of node-line SMs ($k_0 \gg |k_F|$), the RKKY interaction in the transverse plane oscillate more rapidly than the case in the vertical direction.

Further, in the intrinsic case, the RKKY interaction becomes

$$H_{RKKY,\chi}^{\rho} = -\frac{k_0}{4\pi^3 V_F \rho^4}\left\{\left[\lambda^2 + J^2 + (2J^2 - 4\lambda^2)\sin^2\left(k_0\rho - \frac{\pi}{4}\right)\right]\vec{S}_1 \cdot \vec{S}_2 - 6(J^2 - \lambda^2)\sin^2\left(k_0\rho - \frac{\pi}{4}\right)S_1^x S_2^x\right\},$$

$$H_{RKKY}^{\rho} = -\frac{k_0 J^2}{4\pi^3 V_F \rho^4}\left[1 - 4\sin^2\left(k_0\rho - \frac{\pi}{4}\right)\right]\vec{S}_1 \cdot \vec{S}_2. \tag{26}$$

It is unusual that, the RKKY interaction of the node-line SMs keeps oscillation with period $\pi k_0^{-1}$ and decays as $1/\rho^4$ in the transverse plane, distinctly different from $1/R^5$ for the intrinsic Dirac and Weyl SMs [32-33]. Consequently, either the FM or AFM coupling of two magnetic impurities depends on the distance between impurities. When $\lambda = J$, the RKKY exchange coupling between two magnetic impurities is always FM in the node-line SMs with chirality.

**C. Possible application in realistic NLSM: ZrSiSe**

A number of materials, such as PbTaSe$_2$ [22], TlTaSe$_2$ [23], and ZrSiM [21, 24-26] (M =S, Se, Te) family, have been predicted to be NLSM. However, the much prominent material is ZrSiM due to the quite large energy window of the extended linear band-crossings [21, 25] along lines/loops in the Brillouin zone. For instance, near the energy region of 2 eV above and below E$_F$, the ZrSiSe bands are almost linear along the directions crossing center of nodal ring [26]. Hence, our model Hamiltonian of NLSMs with linear dispersion in the nodal-line plane may be applicable to ZrSiSe. Considering the anisotropy in realistic ZrSiSe material, the effective $k \cdot p$ Hamiltonian of NLSM takes the form $H_0 = \left[V_\rho(k_\rho - k_0)\sigma_x + V_z k_z \sigma_y\right]$. The Eq.(1) is identical to this Hamiltonian through a scale transformation $(V_F, k_z, z) \rightarrow (V_\rho, k_z V_z/V_\rho, z V_\rho/V_z)$, so we can easily obtain the RKKY interaction of anisotropic NLSM by multiplying $(V_\rho/V_z)^2$ on the preceding expressions of RKKY interaction and recovering the scales.

From the ARPES data of ZrSiSe [26], we can estimate band parameters $V_\rho \approx 4.5\ eV \cdot Å$, $k_0 \approx 0.4\ Å^{-1}$, and $\varepsilon_F \approx 0.34\ eV$. The value of $V_z$ is uncertain but we assume $V_z \approx 3\ eV \cdot Å$. J and $\lambda$ is around $100 \sim 200\ eV \cdot Å^3$ [39, 40]. For simplicity, we assume $\lambda = J = 100\ eV \cdot Å^3$ and then only the Heisenberg term is left in the system of ZrSiSe. We plot the possible range functions of the RKKY interaction of ZrSiSe as a function of the distance between two magnetic impurities for the cases with and without the chirality in Fig. 3.

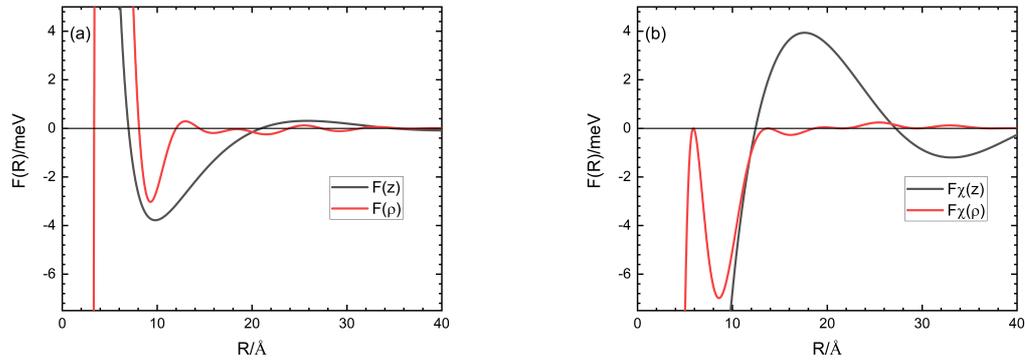

Fig. 3 The range function of the RKKY interaction of ZrSiSe as a function of the distance between two magnetic impurities. The (a) is the case without chirality and the (b) is the case with chirality. The black lines are the impurities aligned along the direction perpendicular to nodal-line plane. The red lines are the impurities aligned in the nodal-line plane.

From Fig. 3, we can see that the magnitude of the strength of the RKKY interaction of ZrSiSe is about several meV around the distance of 10 Å. And the range functions oscillate more rapidly with the smaller period, about 7 Å, for the impurities aligned in the nodal-line plane, which is the direct consequence of large nodal-line radius of ZrSiSe. Furthermore, if the system of ZrSiSe has the chirality, due to the $1/R^2$ decaying law, the RKKY interaction is prominently strongest for the impurities aligned along the direction perpendicular to nodal-line plane. Comparing with the vertical direction, we find whether the system has the chirality or not, the strength of the RKKY interaction decay to very weak when the separation is larger than 12 Å for the impurities aligned in the nodal-line plane.

## IV. DISSCUSION AND CONCLUSION

Further, we briefly discuss the case of NLSMs with quadratic dispersion [36] in the nodal-line plane. Derivation procedure and specific expressions of RKKY interaction are presented in the Appendix A. Due to the similar expressions of the Green's function within the large ring approximation, the RKKY interaction has no essential difference in oscillation and decay features. The major difference is the power law dependence of RKKY amplitude on the node-line radius $k_0$. For the impurities aligned along the vertical direction, the amplitude is independent of $k_0$. For the impurities aligned in the transverse plane, the power-law dependence of amplitude on $k_0$ amounts to $k_0^2$ at short range and $k_0^1$ at long range.

In summary, we have investigated the RKKY interaction between magnetic impurities in NLSMs with and without the chirality. In the large ring limit, we obtained approximately analytical formula of the RKKY interaction for two representative impurity alignments. The NLSMs with linear and quadratic dispersions in the nodal-line plane are both discussed and there is no essential difference in solutions. We found that, in contrast to the previous studies of topological semimetals with the point nodes, distinct torus FS of nodal-line SMs has significant influences on the RKKY interaction, leading to special results. The oscillation forms a beating pattern and depend on the direction between two magnetic impurities. And the decay features are unique in two representative directions, absolutely different from $1/R^5$ ($1/R^3$) decaying law for zero (finite) Fermi energy in the Dirac/Weyl SMs. The most prominent is a typical two-dimensional decay feature for magnetic impurities embedded deeply inside nodal-line SMs bulk and aligned along the direction perpendicular to nodal-line plane. The possible application of our results in realistic NLSM ZrSiSe is discussed. Our findings may be helpful for application of node-line SMs in spintronics and provide a potential way for developing quantum topological devices based on the NLSMs.

*Note added*. When we finish this work we aware that a paper published in Phys. Rev. B in Ref. [42] also concerned similar problem and obtained numerical results, in contrast, we present approximately accurately analytical results and possible application.

***Acknowledgements:*** The supports from the NSFC of China under Grant Nos. 11534010, 11974354 and 11774350 are acknowledged. Numerical calculations were performed at the Center for Computational Science of CASHIPS.

**Appendix A: The RKKY interaction in nodal-line SMs with quadratic dispersions in the nodal-line plane**

For the case of quadratic dispersions in the nodal-line plane, the Hamiltonians of nodal-line SMs [36] is

$$H_0 = \chi[A(k_r^2 - C)\sigma_x + Bk_z\sigma_y], \tag{A1}$$

where $A = \frac{1}{2m_r}$, $B = V_z$, $C = k_0^2$.

In this system, the RKKY interactions have the same forms in Eq. (8) and Eq. (9). The only difference is the partial Green's functions

$$G_i(\vec{R},\varepsilon) = \frac{1}{(2\pi)^2}\frac{A}{B}(\int_0^C dS \int_0^\pi d\phi + \int_C^\infty dS \int_0^{arccos(-C/S)} d\phi)\frac{J_0[\sqrt{(S\cos\phi+C)}r]}{(\varepsilon+i\eta)^2 - A^2S^2}SD_i, (i = 0,1,2) \tag{A2}$$

where $D_1 = \varepsilon \cos(S\tilde{z}\sin\phi)$, $D_2 = A\cos(S\tilde{z}\sin\phi)S\cos\phi$, $D_3 = iA\sin(S\tilde{z}\sin\phi)S\sin\phi$, $\tilde{z} = (A/B)z$.

Within the large ring approximation ($k_0^2/2m_r \gg |\varepsilon|$), the $G_i(\vec{R},\varepsilon)$ can be simplified to

$$g_i(\vec{R},\varepsilon) = \frac{1}{(2\pi)^2}\frac{A}{B}\int_0^\infty dS \int_0^\pi d\phi \frac{J_0[\sqrt{C}(1+\frac{S}{2C}\cos\phi)r]}{(\varepsilon+i\eta)^2 - A^2S^2}SD_i, (i = 0,1,2) \tag{A3}$$

which have the similar forms in the Eq. (11). Thus repeating the same derivation procedure in the Sec. III, we obtain the RKKY interactions for impurities aligned along the vertical direction

$$H_{RKKY,\chi}^z = [\lambda^2 F_1(z) + J^2 F_2(z)](S_1^x S_2^x + S_1^z S_2^z) + [\lambda^2 F_2(z) + J^2 F_1(z)]S_1^y S_2^y, \tag{A4}$$

where the range functions $F_{1,2}(z)$ are given by

$$F_1(z) = \frac{m_r^2}{4\pi^2}\frac{V_z}{z^3}\cos\left(\frac{2\varepsilon_F}{V_z}z\right),$$

$$F_2(z) = -\frac{m_r^2}{4\pi^2}\frac{V_z}{z^3}\left[2\cos\left(\frac{2\varepsilon_F}{V_z}z\right) + \frac{2\varepsilon_F}{V_z}z\sin\left(\frac{2\varepsilon_F}{V_z}z\right)\right]. \tag{A6}$$

Similarly, the RKKY interactions for impurities aligned in the transverse plane is

$$H_{RKKY,\chi}^\rho = [\lambda^2 F_1(\rho) + J^2 F_2(\rho)](S_1^y S_2^y + S_1^z S_2^z) + [\lambda^2 F_2(\rho) + J^2 F_1(\rho)]S_1^x S_2^x, \tag{A7}$$

where the range functions $F_{1,2}(\rho)$ are given by

$$F_1(\rho) = \frac{2}{\pi k_0 \rho}\left[\cos^2\left(k_0\rho - \frac{\pi}{4}\right)f_0(\rho) - \sin^2\left(k_0\rho - \frac{\pi}{4}\right)f_2(\rho)\right],$$

$$F_2(\rho) = \frac{2}{\pi k_0 \rho}\left[\cos^2\left(k_0\rho - \frac{\pi}{4}\right)f_0(\rho) + \sin^2\left(k_0\rho - \frac{\pi}{4}\right)f_2(\rho)\right],$$

with

$$f_0(\rho) = -\frac{1}{8\pi^2 m_r V_z^2}\frac{k_0^3}{\rho^3}\left[\cos\left(\frac{2m_r\varepsilon_F}{k_0}\rho\right) + \frac{2m_r\varepsilon_F}{k_0}\rho\sin\left(\frac{2m_r\varepsilon_F}{k_0}\rho\right)\right],$$

$$f_2(\rho) = -\frac{1}{8\pi^2 m_r V_z^2}\frac{k_0^3}{\rho^3}\left[3\cos\left(\frac{2m_r\varepsilon_F}{k_0}\rho\right) + \frac{2m_r\varepsilon_F}{k_0}\rho\sin\left(\frac{2m_r\varepsilon_F}{k_0}\rho\right)\right]. \quad (A8)$$

**References**

[1] C.-Z. Chang, J. Zhang, X. Feng, J. Shen, Z. Zhang, M. Guo, K. Li, Y. Ou, P. Wei, L.-L. Wang, Z.-Q. Ji, Y. Feng, S. Ji, X. Chen, J. Jia, X. Dai, Z. Fang, S.-C. Zhang, K. He, Y. Wang, L. Lu, X.-C. Ma, and Q.-K. Xue, Science **340**, 167 (2013).

[2] R. Yu, W. Zhang, H.-J. Zhang, S.-C. Zhang, X. Dai, Z. Fang, Science **329**, 61 (2010).

[3] H. Zhang, C. X. Liu, X. L. Qi, X. Dai, Z. Fang, and S.-C. Zhang, Nat. Phys. **5**, 438 (2009).

[4] H. Lin, R. S. Markiewicz, L. A. Wray, L. Fu, M. Z. Hasan, and A. Bansil, Phys. Rev. Lett. **105**, 036404 (2010).

[5] L. Fu and C. L. Kane, Phys. Rev. B **76**, 045302 (2007).

[6] M. A. Ruderman and C. Kittel, Phys. Rev. **96**, 99 (1954).

[7] K. Yosida, Phys. Rev. **106**, 893 (1957).

[8] T. Kasuya, Prog. Theor. Phys. **16**, 45 (1956).

[9] T. Yamamoto, and F. J. Ohkawa, J. Phys. Soc. Jpn. **57**, 3562 (1988).

[10] Y. Ōnuki, R. Settai, K. Sugiyama, T. Takeuchi, T. C. Kobayashi, Y. Haga, and E. Yamamoto, J. Phys. Soc. Jpn. **73**, 769 (2004).

[11] E. Z. Meilikhov, Phys. Rev. B **75**, 045204 (2007).

[12] D. J. Priour, Jr., E. H. Hwang, and S. Das Sarma, Phys. Rev. Lett. **92**, 117201 (2004).

[13] Q. Liu, C.-X. Liu, C. Xu, X.-L. Qi, and S.-C. Zhang, Phys. Rev. Lett. **102**, 156603 (2009).

[14] S. M. Young, S. Zaheer, J. C. Y. Teo, C. L. Kane, E. J. Mele, and A. M. Rappe, Phys. Rev. Lett. **108**, 140405 (2012).

[15] Z. K. Liu, B. Zhou, Y. Zhang, Z. J. Wang, H. M. Weng, D. Prabhakaran, S.-K. Mo, Z. X. Shen, Z. Fang, X. Dai, Z. Hussain, and Y. L. Chen, Science **343**, 864 (2014).

[16] Z. K. Liu, J. Jiang, B. Zhou, Z. J. Wang, Y. Zhang, H. M.Weng, D. Prabhakaran, S.-K. Mo, H. Peng, P. Dudin, T. Kim, M. Hoesch, Z. Fang, X. Dai, Z. X. Shen, D. L. Feng, Z. Hussain, and Y. L. Chen, Nat. Mater. **13**, 677 (2014).

[17] M. Neupane, S.-Y. Xu, R. Sankar, N. Alidoust, G. Bian, C. Liu, I. Belopolski, T.-R. Chang, H.-T. Jeng, H. Lin, A. Bansil, F. Chou, and M. Z. Hasan, Nat. Commun. **5**, 3786 (2014).

[18] Y. Zhang, Y.-W. Tan, H. L. Stormer, and P. Kim, Nature **438**, 201 (2005).

[19] B. Q. Lv, N. Xu, H. M. Weng, J. Z. Ma, P. Richard, X. C. Huang, L. X. Zhao, G. F. Chen, C. E. Matt, F. Bisti, V. N. Strocov, J. Mesot, Z. Fang, X. Dai, T. Qian, M. Shi, and H. Ding, Nat. Phys. **11**, 724 (2015).

[20] X. Wan, A. M. Turner, A. Vishwanath, and S. Y. Savrasov, Phys. Rev. B **83**, 205101 (2011).

[21] Y. Shao, A. N. Rudenko, J. Hu, Z. Sun, Y. Zhu, S. Moon, A. J. Millis, S. Yuan, A. I. Lichtenstein, D. Smirnov, Z. Q. Mao, M. I. Katsnelson, and D. N. Basov, Nat. Phys. **16**, 636 (2020).

[22] G. Bian, T.-R. Chang, R. Sankar, S.-Y. Xu, H. Zheng, T. Neupert, C.-K. Chiu, S.-M. Huang, G. Chang, I. Belopolski, D. S. Sanchez, M. Neupane, N. Alidoust, C. Liu, B. Wang, C.-C. Lee, H.-T. Jeng, C. Zhang, Z. Yuan, S. Jia, A. Bansil, F. Chou, H. Lin, and M. Z. Hasan, Nat. Commun. **7**, 10556 (2016).

[23] G. Bian, T.-R. Chang, H. Zheng, S. Velury, S.-Y. Xu, T. Neupert, C.-K. Chiu, S.-M. Huang, D. S. Sanchez, I. Belopolski, N. Alidoust, P.-J. Chen, G. Chang, A. Bansil, H.-T. Jeng, H. Lin, and M. Z. Hasan, Phys. Rev. B **93**, 121113 (2016).

[24] A. N. Rudenko, E. A. Stepanov, A. I. Lichtenstein, and M. I. Katsnelson, Phys. Rev. Lett. **120**, 216401 (2018).

[25] F. C. Chen, Y. Fei, S. J. Li, Q. Wang, X. Luo, J. Yan, W. J. Lu, P. Tong, W. H. Song, X. B. Zhu, L. Zhang, H. B. Zhou, F. W. Zheng, P. Zhang, A. L. Lichtenstein, M. I. Katsnelson, Y. Yin, N. Hao, and Y. P. Sun, Phys. Rev. Lett. **124**, 236601 (2020).

[26] G. Gatti, A. Crepaldi, M. Puppin, N. Tancogne-Dejean, L. Xian, U. De Giovannini, S. Roth, S. Polishchuk, P. Bugnon, A. Magrez, H. Berger, F. Frassetto, L. Poletto, L. Moreschini, S. Moser, A. Bostwick, E. Rotenberg, A. Rubio, M. Chergui, and M. Grioni, Phys. Rev. Lett. **125**, 076401 (2020).

[27] G. P. Mikitik and Yu. V. Sharlai, Phys. Rev. B **94**, 195123 (2016).

[28] L. Brey, H. A. Fertig, and S. D. Sarma, Phys. Rev. Lett. **99,** 116802 (2007).

[29] S. Saremi, Phys. Rev. B **76**, 184430 (2007).

[30] E. Kogan, Phys. Rev. B **84**, 115119 (2011).

[31] B. Fischer and M. W. Klein, Phys. Rev. B **11**, 2025 (1975).

[32] H.-R. Chang, J. Zhou, S.-X. Wang, W.-Y. Shan, and D. Xiao, Phys. Rev. B **92**, 241103 (2015).

[33] M. V. Hosseini, and M. Askari, Phys. Rev. B **92**, 224435 (2015).

[34] H.-J. Duan, S.-H. Zheng, P.-H. Fu, R.-Q. Wang, J.-F. Liu, G.-H. Wang, and M. Yang, New J. Phys. **20**, 103008 (2018).

[35] J.-W. Rhim, and Y. B. Kim, New J. Phys. **18**, 043010 (2016).

[36] H. Yang, R. Moessner, and L.-K. Lim, Phys. Rev. B **97**, 165118 (2018).

[37] Y. Sun and A. Wang, J. Phys.: Condens. Matter **29**, 435306 (2017).

[38] H. Imamura, Phys. Rev. B **69**, 121303 (2004).

[39] J. S. Dyck, P. Ha´jek, P. Losˇt'a´k, and C. Uher, Phys. Rev. B, **65**, 115212 (2002).

[40] F. Matsukura, H. Ohno, A. Shen, and Y. Sugawara, Phys. Rev. B **57**, R2037 (1998).

[41] P.-L. Gong, D.-Y. Liu, K.-S. Yang, Z.-J. Xiang, X.-H. Chen, Z. Zeng, S.-Q. Shen, and L.-J. Zou, Phys. Rev. B **93**, 195434 (2016)

[42] H.-J. Duan, S.-H. Zheng, Y.-Y. Yang, C.-Y. Zhu, M.-X. Deng, M. Yang, and R.-Qiang Wang, Phys. Rev. B 102, 165110 (2020).